\documentstyle[epsfig]{mn}
\DeclareMathVersion{bold}

\title[Spherical wavelets and non-Gaussianity in the COBE-DMR data]
{Testing the Gaussianity of the COBE-DMR data with spherical wavelets}
\author[R.B.~Barreiro et al.]{R.B.~Barreiro$^1$, M.P.~Hobson$^1$,
A.N.~Lasenby$^1$, A.J.~Banday$^2$, K.M.~G\'orski$^{3,4}$, 
\newauthor and G.~Hinshaw$^5$\\
$^1$ Astrophysics Group, Cavendish Laboratory, Madingley Road,
Cambridge CB3 0HE, UK \\
$^2$ Max-Planck Institut fuer Astrophysik (MPA), Karl-Schwarzschild
Str.1, D-85740, Garching, Germany \\
$^3$ European Southern Observatory (ESO), Karl-Schwarzschild
Str.2, D-85740, Garching, Germany \\
$^4$ Warsaw University Observatory, Aleje Ujazdowskie 4, 00-478
Warszawa, Poland \\
$^5$ NASA/GSFC, Greenbelt, MD 20771, USA }
\date{Accepted ???. Received ???; in original form ???}

\begin{document}
\maketitle
\begin{abstract}
We investigate the Gaussianity of the 4-year COBE-DMR data (in HEALPix
pixelisation) using an analysis based on spherical Haar wavelets. 
We use all the pixels lying outside the Galactic cut and
compute the skewness, kurtosis and scale-scale correlation spectra for
the wavelet coefficients at each scale. 
We also take into account the sensitivity of the
method to the orientation of the input signal.
We find a detection of non-Gaussianity at $> 99$ per cent level in just one of
our statistics. Taking
into account the total number of statistics computed, we estimate 
that the probability of obtaining such a
detection by chance for an underlying Gaussian field is 0.69.
Therefore, we conclude that the spherical wavelet 
technique shows no strong evidence of non-Gaussianity in the COBE-DMR data.
\end{abstract}

\begin{keywords}
methods: data analysis-cosmic microwave background.
\end{keywords}

\section{Introduction}

The cosmic microwave background (CMB) provides a unique tool for
investigating the formation of structure in the Universe. 
In particular, studying the Gaussianity of the CMB temperature fluctuations
allows us to distinguish between two competing
theories of structure formation: the standard inflationary model that predicts
Gaussian fluctuations and topological defects that give rise to
non-Gaussian signatures in the CMB. In order to test the 
Gaussianity of the CMB, a large number of
methods have already been proposed in the literature (see e.g.
Barreiro 2000 and references therein). In particular, many of them
have been applied to the 4-year COBE-DMR data, although most have
not yielded any detection of non-Gaussianity (see e.g. Mukherjee, Hobson \&
Lasenby 2000).

Nevertheless, three apparently robust detections of non-Gaussianity 
in the 4-year COBE-DMR data have recently been reported. 
Ferreira, Magueijo \& G\'orski (1998) studied the
distribution of an estimator for the normalised bispectrum, finding
that Gaussianity is ruled out at the $98$ per cent confidence level with the
non-Gaussian signal mainly concentrated on the multipole $l=16$.
Although this non-Gaussian signal is certainly present in the data,
Banday, Zaroubi \& G\'orski (1999) have shown in a recent work that
this non-Gaussian signal is most probably due to a systematic
artefact. However, Magueijo (1999) has found a new non-Gaussian signal
above the $97$ per cent confidence level using an extended bispectrum
analysis, which is present even when the artefacts
found by Banday et al. are removed.

Complementary to the above bispectrum analysis, 
Pando, Valls-Gabaud \& Fang (1998) applied a
technique based on the discrete wavelet transform to Face 0 and
Face 5 of the QuadCube pixelisation of COBE-DMR data.
On computing the scale-scale correlations of the wavelet coefficients in
certain domains of the wavelet transform, they found a
significant non-Gaussian signal at the 99 per cent confidence level in Face 0
corresponding to scales of $11-22$ degrees. However, no significant
deviation from Gaussianity was found using the skewness
and kurtosis of the wavelet coefficients at each of the considered domains.
Mukherjee, Hobson \& Lasenby (2000) (hereinafter MHL) 
have recently revised the previous work, taking into account that a
large number of the computed statistics show no evidence of 
non-Gaussianity, and
pointing out that the results depend critically on the orientation of
the data. They find that Gaussianity can only be ruled out at the
41 per cent confidence level in the DSMB data and at the 72 per cent
level in the 53+90 GHz coadded data and therefore that this analysis does
not provide strong evidence for non-Gaussianity in the COBE-DMR data.

The above wavelet analyses were performed 
by applying planar (Daubechies) wavelets to Face 0 and Face 5 of the
COBE-DMR QuadCube pixelisation in Galactic coordinates (i.e. the faces
centred on the North and South Galactic poles respectively).
This procedure thus uses only one-third of the COBE-DMR data and,
furthermore, can lead to distortions of the CMB fluctuations when
moving from the sphere to the planar faces of the QuadCube. 
Therefore, in this paper, we use
orthogonal spherical Haar 
wavelets (SHW) (Girardi \& Sweldens 1995, Sweldens 1995), which are
better suited
to analysing data over large regions of the sky (such as the COBE-DMR maps).
These wavelets were introduced as a generalisation of
planar Haar wavelets to more general spaces than $R^n$. Indeed, Tenorio et
al. (1999) have used SHW as a tool for studying the spatial structure, 
denoising and compression of CMB maps.
We use SHW to perform a similar analysis to 
those of Pando et al and MHL, based on the skewness, kurtosis and
scale-scale correlation of the wavelet coefficients, in order to
search for evidence of non-Gaussianity in the 4-year COBE-DMR data.
A hierarchical pixelisation scheme which is particularly well-suited to
the application of such a wavelet decomposition is HEALPix (G\'orski,
Hivon \& Wandelt 1999).
Therefore, we apply the former analysis to the COBE-DMR data in HEALPix
pixelisation (Banday et al. 2000). We also include a Galactic cut
derived by a simple propagation of the customised Galactic cut of
Banday et al. (1997) and Bennett et al. (1996) to the HEALPix
pixelisation.
Owing to the characteristics of the SHW transform, it is straightforward 
to use {\em all} the pixels lying outside the Galactic cut, which
constitutes approximately two-thirds of the total number of COBE-DMR pixels.


\section{The wavelet analysis}
\subsection{Spherical Haar Wavelets}

Orthogonal SHW are an example of the so-called `second generation
wavelets' (Sweldens 1995,  Schroeder \& Sweldens 1995). These 
wavelets are not dilations and translations of a given function
and can therefore be adapted to more general spaces than $R^n$ but, at
the same time, they still enjoy all the useful properties of planar
wavelets, such as good space-frequency localisation and a fast
transform algorithm.

The temperature field can be decomposed into the SHW basis functions:
\begin{eqnarray}
\frac{\Delta T}{T}(x_i)&=&\sum_{l=0}^{n_{j_0}-1}
\lambda_{j_0,l}\varphi_{j_0,l}(x_i) \nonumber \\
& & +
\sum_m\sum_{j=j_0}^{J-1}\sum_{l=0}^{n_j-1}\gamma_{m,j,l}\psi_{m,j,l}(x_i)
\label{shweqn}
\end{eqnarray} 
where $\lambda_{j_0,l}$ and $\gamma_{m,j,l}$ are the approximation and
detail wavelet coefficients respectively. The first term in (\ref{shweqn})
corresponds to a smoother image of the original map, 
whereas the detail coefficients encode the differences between the
smoothed map and the original.
The index $j$ runs over the different scales, with $J$ being the resolution of 
the original map and $j_0$ the coarsest resolution
considered. $n_j$ is the number of pixels at resolution $j$.
Finally, the index $m$ corresponds to the number of different wavelet
functions at each scale required in order to form a complete
orthogonal basis set.
In particular, for a square partitioning, such as that in the HEALPix
pixelisation,
we require three different wavelet basis functions at each scale $j$,
and therefore we have three
different kind of wavelet coefficients
$\gamma_{1,j,l}$, $\gamma_{2,j,l}$ and $\gamma_{3,j,l}$.
A more detailed description of the SHW transform is given
in Appendix~\ref{apendice}.

As pointed out by MHL, we note that
the asymmetry of orthogonal wavelet basis functions means that
the wavelet decomposition is sensitive to the orientation of
the input signal. 
Thus any statistics based on the corresponding wavelet coefficients
are also sensitive to the orientation of the analysed signal.

\subsection{Application to COBE-DMR data}

In this paper, we analyse the 4-year COBE-DMR data in HEALPix pixelisation. 
A detailed description of the process of map-making is given by
Banday et al. (2000). 
HEALPix is an equal area, iso-latitude and hierarchical
pixelisation of the sphere. The base-resolution comprises
twelve pixels in three rings around the poles and equator
(see Fig.~\ref{cobemap}). The
resolution level of the grid is expressed by the parameter $N_{\rm side}$,
that indicates the number of divisions along the side of the
base-resolution pixel that is needed to reach a desired
high-resolution partition. For convenience we will use instead the
index $j$ to refer to the scale; this relates to $N_{\rm side}$ as
$N_{\rm side}=2^{j-1}$. 

In the case of the COBE-DMR maps, the total number of pixels is 49152 
(i.e. $J=7$, or
equivalently $N_{\rm side}=64$), with a pixel linear size of $\simeq 55'$.
Therefore, each resolution level $j$ corresponds to a scale 
$55'\times 2^{J-j}$, containing $n_{j}=12\times 4^{j-1}$
pixels of that size.
\begin{figure}
\bigskip
\caption{In the top figure, the coadded 53+90 GHz COBE-DMR map in
HEALPix pixelisation at resolution $J=7$ with the customised Galactic
cut is plotted.
The bottom map shows the twelve pixels of the base-resolution of
HEALPix ($j=1$) and how they are affected by such a cut.}
\label{cobemap}
\end{figure}

We have used in our analysis the coadded 53A, 53B, 90A and 90B map 
(each pixel weighted according to the inverse of its noise variance)
with the customised Galactic cut, which is plotted in
Fig.~\ref{cobemap} (Banday et al. 2000, Banday et al. 1997, 
Bennett et al. 1996). As mentioned above, 
we are thus using approximately two-thirds of the
data as opposed to previous analyses that kept only one-third (Face 0 and
Face 5) of the QuadCube COBE-DMR data. In addition, we avoid any possible
projection effects in going from the sphere to the cube.

\subsection{The non-Gaussianity test}

In our wavelet analysis we have considered separately the detail
coefficients corresponding to $m = 1,2,3$ at each scale $j$.
For each value of $j$ and $m$, we use the corresponding detail
coefficients $\gamma_{m,j,l}$ to estimate the skewness $\hat{S}$ and (excess)
kurtosis $\hat{K}$ of the parent distribution from which the coefficients 
were drawn. We therefore obtain the skewness $\hat{S}(j,m)$ and
kurtosis $\hat{K}(j,m)$ spectra for the image.
At each value $(j,m)$, the skewness and kurtosis of the parent
distribution of the wavelet coefficients is given by
\begin{equation}
S=\kappa_3/\kappa_2^{3/2}
\end{equation}
\begin{equation}
K=\kappa_4/\kappa_2^{2}
\end{equation}
where $\kappa_n$ is the $n$th cumulant of the distribution.
Following Hobson, Jones \& Lasenby (1999) we use $k$-statistics
(see Kenney \& Keeping 1954; Stuart \& Ord 1994) to obtain unbiased 
estimators of the cumulants, which are then used to estimate
the skewness $\hat{S}$ and (excess) kurtosis $\hat{K}$. 

In addition to the skewness and kurtosis spectra, we measure the
scale-scale correlation between the wavelet coefficients at different
scales using the following estimator:
\begin{equation}
\hat{C}^p(j,m)=\frac{n_{j+1}\sum_{l=0}^{n_{j+1}-1}\gamma^p_{m,j,[l/4]}
\gamma^p_{m,j+1,l}}{\sum_{l=0}^{n_{j+1}-1}\gamma^p_{m,j,[l/4]}
\sum_{l=0}^{n_{j+1}-1}\gamma^p_{m,j+1,l}}
\end{equation}
where [~] denotes the integer part and $n_j=12\times 4^{j-1}$ is the
number of pixels at the resolution level $j$. The scale-scale correlation
coefficient $\hat{C}^p(j,m)$ measures the correlation between each
type of detail coefficient $\gamma$ (i.e. for $m=1,2,3$) 
in the two consecutive scales $j$ and $j+1$. 
We restrict our analysis to the case $p=2$.

In our non-Gaussianity test, we first obtain the skewness, kurtosis
and scale-scale correlation spectra for the coadded 53+90 GHz
COBE-DMR map. Those wavelet coefficients sensitive to pixels inside the
Galactic cut are not used in the computations.
We then generate 10000 realisations of CMB all-sky maps, smoothed with
a $7^\circ$ Gaussian beam, and add levels of Gaussian noise 
corresponding to the
COBE-DMR data. The CMB maps are drawn from an inflationary/CDM model with
parameters $\Omega_m=1$, $\Omega_\Lambda=0$, $h=0.5$, $n=1$ and
$Q_{rms-ps}=18\mu K$, but we do not expect our analysis to be very
sensitive to a difference choice of parameters (see MHL).
We then compute our estimators for each of the CDM simulations to
obtain approximate probability distributions for the $\hat{S}(j,m)$,
$\hat{K}(j,m)$ and $\hat{C}^2(j,m)$ statistics for a 
CMB signal derived from the chosen model.
By comparing these distributions with the values obtained from the
COBE-DMR data, we can obtain the probability that our data are derived
from a Gaussian distribution characterised by the chosen power spectrum.

For each kind of detail
coefficient ($m=1,2,3$), we calculate the skewness and kurtosis
at five different scales, and also calculate four 
scale-scale correlation coefficients. Therefore we 
compute a total of 42 different statistics, which must be taken into
account when assigning a statistical significance to a given
detection. In addition, as we have already discussed, this wavelet
decomposition is sensitive to the orientation of the data. 
Therefore, we perform the entire test for three different orientations
of the COBE-DMR map.

\section{Results}
\begin{figure*}
\centerline{\epsfig{
file=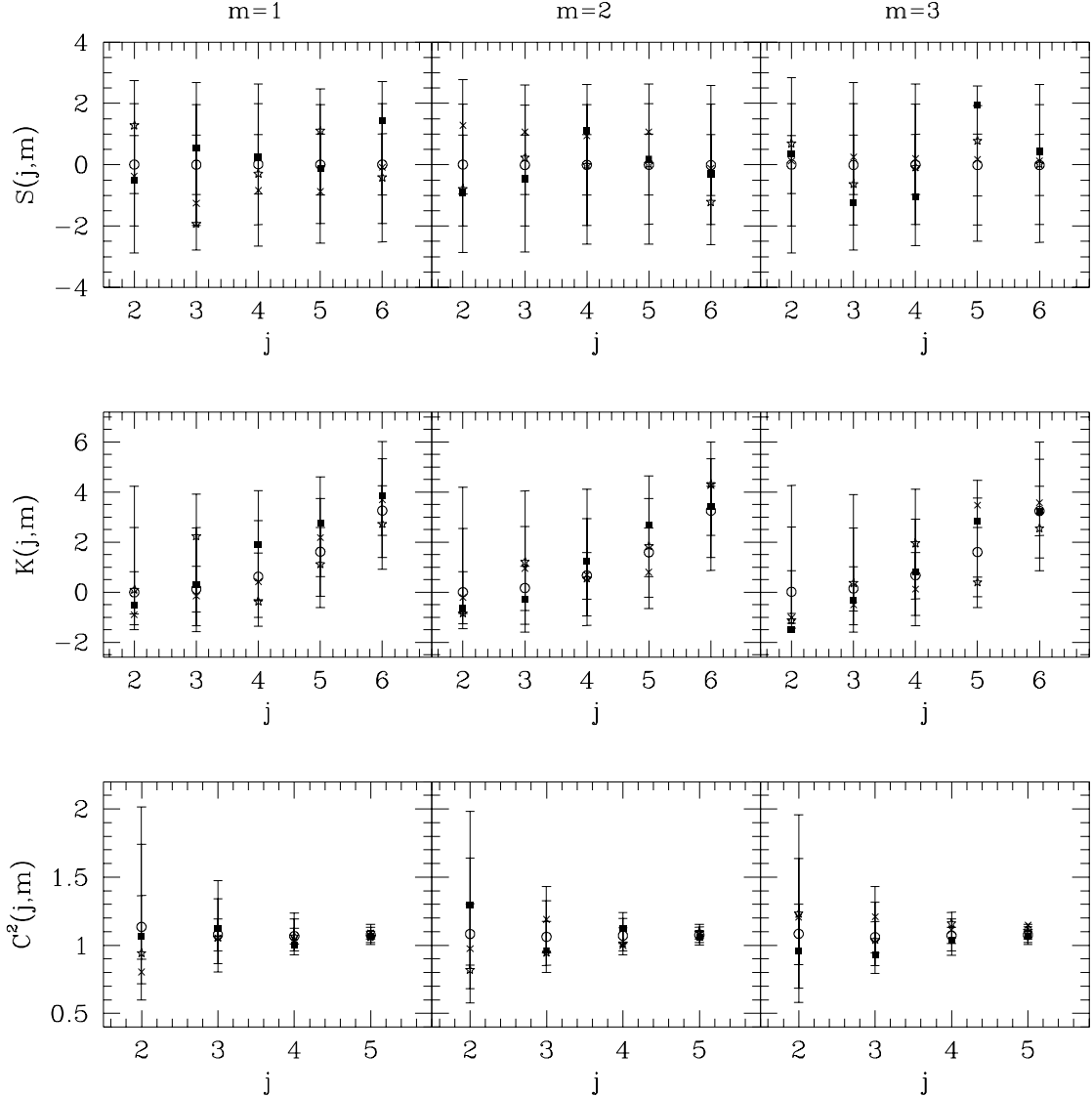,
width=16cm
}}
\caption{The skewness, kurtosis and scale-scale correlation spectra
for the 53+90 GHz COBE-DMR map
are plotted for each type of detail coefficient $m=1$ (first
column), $m=2$ (second column) and $m=3$ (third column).
Solid squares, crosses and stars correspond respectively to the the
data rotated 0, 30 and 60 degrees around an axis passing through the
North and South Galactic poles
with respect to the input signal (orientations A, B and C).
The error bars denote the 68, 95 and 99 per cent limits of the
corresponding distribution. For convenience, $\hat{S}(j,m)$ and
$\hat{K}(j,m)$ have been normalised at each value of $(j,m)$ such that
the variance of the distribution obtained from the 10000 CDM
realisations is equal to unity}
\label{spectra}
\end{figure*}
\begin{figure*}
\centerline{\epsfig{
file=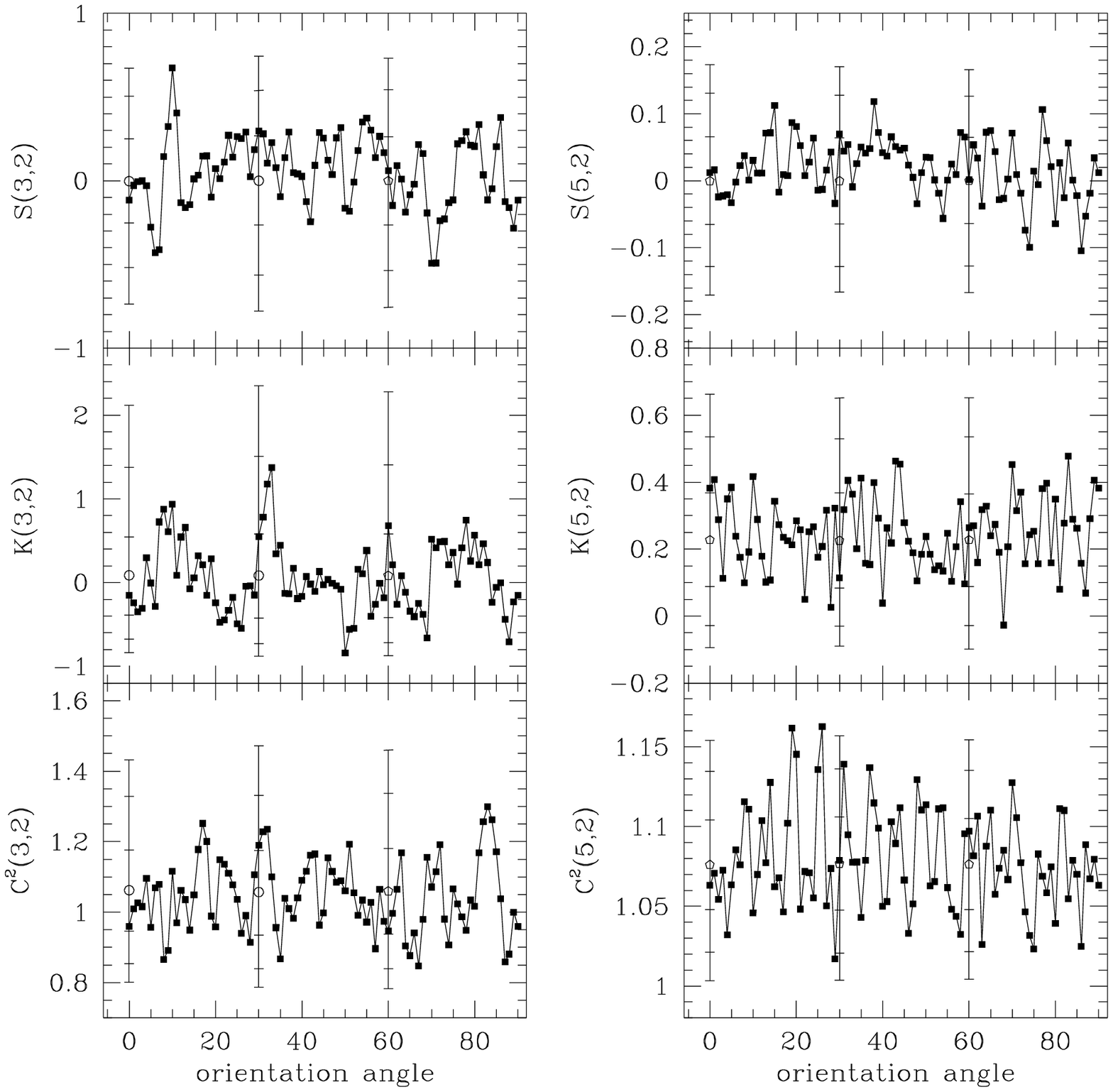,
width=16cm
}}
\caption{The $\hat{S}(j,m)$,$\hat{K}(j,m)$ and $\hat{C^2}(j,m)$
spectra are shown for values of $j=3,5$ and $m=2$
versus the angle that the COBE-DMR data have been rotated around an axis
passing through the North and South Galactic poles
with respect to the original signal. 
The average value and the $68$, $95$ and $99$ per cent confidence levels
obtained from the corresponding Gaussian distributions are also
plotted for values of the orientation angle equal to 0, 30 and 60
degrees (orientations A, B and C, respectively). Note that the skewness
and kurtosis spectra have {\em not} been normalised as in Fig.~\ref{spectra}}
\label{orientation}
\end{figure*}

The computed $\hat{S}(j,m)$, $\hat{K}(j,m)$ and $\hat{C}^2(j,m)$
spectra are plotted in Fig.~\ref{spectra} for three different
orientations. We rotate the 53+90 COBE-DMR map around an axis passing
through the North and South Galactic poles. Due to
the characteristics of the HEALPix pixelisation, a rotation in this
direction of $90$ degrees simply shifts the twelve base-pixels 
into each other, and thus recovers the original result.
Therefore, we show the results computed from the COBE-DMR data for a
rotation of 0 (orientation A, solid squares), 30 (orientation B,
crosses) and 60 (orientation C, stars) degrees with respect to the 
original orientation of the 53+90 
COBE-DMR map. The open circle corresponds to the average over 10000 CDM
simulations generated as explained before and the error bars indicate
the 68, 95 and 99 per cent confidence levels of the distributions for
orientation A. 
Note that for convenience the $K(j,m)$ and $S(j,m)$ spectra have been
normalised at each value of $(j,m)$ such that the variance of the
distribution obtained from the 10000 CDM realisations is equal to unity.
We do not plot the
corresponding error bars for orientations B and C in Fig.~\ref{spectra} 
for the sake of clarity and because the conclusions derived from the plot
remain unchanged. In fact, due to the isotropy of the CMB, 
we expect these error distributions to be independent of the chosen
orientation, although small variations do appear (see
Fig.~\ref{orientation}) due to the presence
of anisotropic noise and as a consequence of having a pixelised map.

We see from Fig.~\ref{spectra} that for orientation A (solid squares) 
all the points of skewness and scale-scale correlation spectra lie
within their respective Gaussian probability distributions. However,
we find a detection of non-Gaussianity at a confidence level $>99$ per cent
for the kurtosis at $j=2$, $m=3$. On the other hand, 
for orientations B (crosses) and C (stars) all the COBE-DMR values are
consistent with being derived from a parent Gaussian distribution.

We must be careful when assessing the significance of the non-Gaussian
detection for orientation A. We must take into account that we have computed a total of 
$3\times 42=126$ different statistics and most of them show
no evidence of non-Gaussianity (see Bromley \& Tegmark
1999). Following MHL, since the different statistics are not independent,
we must use Monte-Carlo simulations to estimate
the probability of having at least one detection of non-Gaussianity
when computing these 126 statistics at a confidence level $>99$ per
cent even in the case of an underlying Gaussian signal. We find
that this occurs in 69 per cent of the cases.
Therefore this analysis does not provide strong evidence of non-Gaussianity,
in agreement with the results obtained by MHL using planar wavelets.

We may wonder, however, if our conclusions are affected
by choosing a different set of orientations for the COBE-DMR maps.
As an illustration of the sensitivity of the computed statistics to
the orientation of the data, Fig.~\ref{orientation} shows 
$\hat{S}(j,m)$,$\hat{K}(j,m)$ and $\hat{C^2}(j,m)$ for $m=2$ and two
different scales versus the
angle that the COBE-DMR data have been rotated around an axis passing
through the North and South Galactic poles with 
respect to the original signal. 
The average value and the $68$, $95$ and $99$ per cent confidence levels
obtained from the corresponding Gaussian distributions are also
plotted for orientations A,B and C. Note that, in this case, the
skewness and kurtosis spectra has {\em not} been normalised as in 
Fig.~\ref{spectra}. It can be seen that the COBE-DMR values of the 
skewness, kurtosis and
scale-scale correlation at each scale oscillate around the mean
obtained from the CDM realisations and with a deviation in agreement
with the confidence limits showed for orientations A,B and C.
We have also checked that, as expected, we 
obtain very similar variations when rotations are performed around 
different directions.
Therefore, although the particular numerical values of the considered
statistics are not rotationally invariant, the general conclusions
concerning non-Gaussianity of the data, at least in our case, do not 
seem to depend on the chosen orientation.

\section{Conclusions}
We presented an analysis of the 4-year COBE-DMR data (in HEALPix pixelisation) 
based on spherical Haar wavelets (SHW) 
in order to look for large scale non-Gaussianity of the CMB. 
This analysis is performed using all the available data lying outside the 
Galactic cut. This constitutes about two-thirds of the pixels, as
compared to just one-third in former works that use only Face 0 and 
Face 5 of the QuadCube pixelisation. We take into account the sensitivity of
the method to the orientation of the original signal and present the 
results for three different orientations of the data. We also find
that the choice of a different set of orientations, at least in our
case, does not alter the general conclusions regarding non-Gaussianity.

We find that the value of the kurtosis for $j=2,m=3$ for one of the
chosen orientations
lies outside the $99$ per cent confidence level derived from 10000
CDM/inflationary realisations, whereas the rest of the statistics show
no evidence of non-Gaussianity. However, since we have a total of 126
different statistics, the probability of finding at least one of them
falling outside the 99 per cent confidence level even in the case of
an underlying Gaussian field is as high as 0.69. Therefore, we
conclude that an analysis based on SHW of the 4-year COBE-DMR data show no 
evidence for non-Gaussianity.

\section*{Acknowledgements}
RBB warmly thanks Luis Tenorio for helpful comments about spherical wavelets.
We thank all the people involved in the HEALPix collaboration
(G\'orski, Hivon \& Wandelt 1999), whose
package has been used extensively in this work.
RBB and MPH acknowledge financial support from the PPARC in the form of a
research grant and an Advanced Fellowship respectively.
ANL acknowledges support from the Royal Society and Leverhulme Trust.

\appendix
\section{Sperical Haar Wavelets for HEALP\lowercase{ix} pixelisation}
\label{apendice}
In this Appendix, we give an heuristic approach to decomposing a map
in HEALPix pixelisation in terms of spherical Haar wavelet
coefficients. More mathematical approaches are given elsewhere 
(Sweldens 1995).


HEALPix is an equal area, iso-latitude and hierarchical pixelisation
of the sphere. The resolution level of the grid is given by the
parameter $j$ (or equivalently $N_{\rm side}$, $N_{\rm side}=2^{j-1}$).
A level $j$ comprises a total of $n_j=12\times 4^{j-1}$ pixels, each
of them with equal area $\mu_j$. 
Each pixel $l$ at resolution $j$, $S_{j,l}$, is divided into four pixels 
$S_{j+1,l_0},...,S_{j+1,l_3}$ at resolution $j+1$.
In particular, the COBE-DMR resolution corresponds to $J=7$.

In order to perform a SHW decomposition of the COBE-DMR data, we require 
one scaling
$\varphi_{j,l}$ and three wavelet functions $\psi_{m,j,l}$ at each 
scale $j$ and position $l$. These are given by
\begin{equation}
\varphi_{j,l}(x)  =  \left\{ \begin{array}{ll}
1, & {\rm if~} x \epsilon S_{j,l}\\
0, & {\rm otherwise} \\
\end{array}
\right.
\end{equation}
\begin{eqnarray}
\psi_{1,j,l}&=&\frac{
\varphi_{j+1,l_0}+\varphi_{j+1,l_2}-
\left(\varphi_{j+1,l_1}+\varphi_{j+1,l_3}\right)}{4\mu_{j+1}}
\\
\psi_{2,j,l}&=&\frac{1}{2\mu_{j+1}}\left(\varphi_{j+1,l_1}-\varphi_{j+1,l_3}\right)\\
\psi_{3,j,l}&=&\frac{1}{2\mu_{j+1}}\left(\varphi_{j+1,l_0}-\varphi_{j+1,l_2}\right)
\end{eqnarray}
where $l_0,l_1,l_2,l_3$ are the four pixels at
resolution level $j+1$ contained within the
pixel $l$ at level $j$, and we have taken into
account that all pixels have equal area at a given resolution. 
The orthogonality of the wavelet functions follows immediately from
their vanishing integral (by construction) and their support.
\begin{figure}
\caption{The COBE-DMR map in HEALPix pixelisation at resolution $J=7$ and
the wavelet coefficients at $j=6$ are shown. Starting from the
coefficients at the highest resolution $J$, the approximation and detail
coefficients at $J-1$ are obtained as linear combinations of the
original image (as shown in Fig~\ref{gcoef}). The process is repeated
over the map formed by the $\lambda_{J-1,l}$ coefficients and so on 
down to the lowest resolution considered $j_0$.}
\label{cobeshw}
\end{figure}
\begin{figure}
\centerline{\epsfig{file=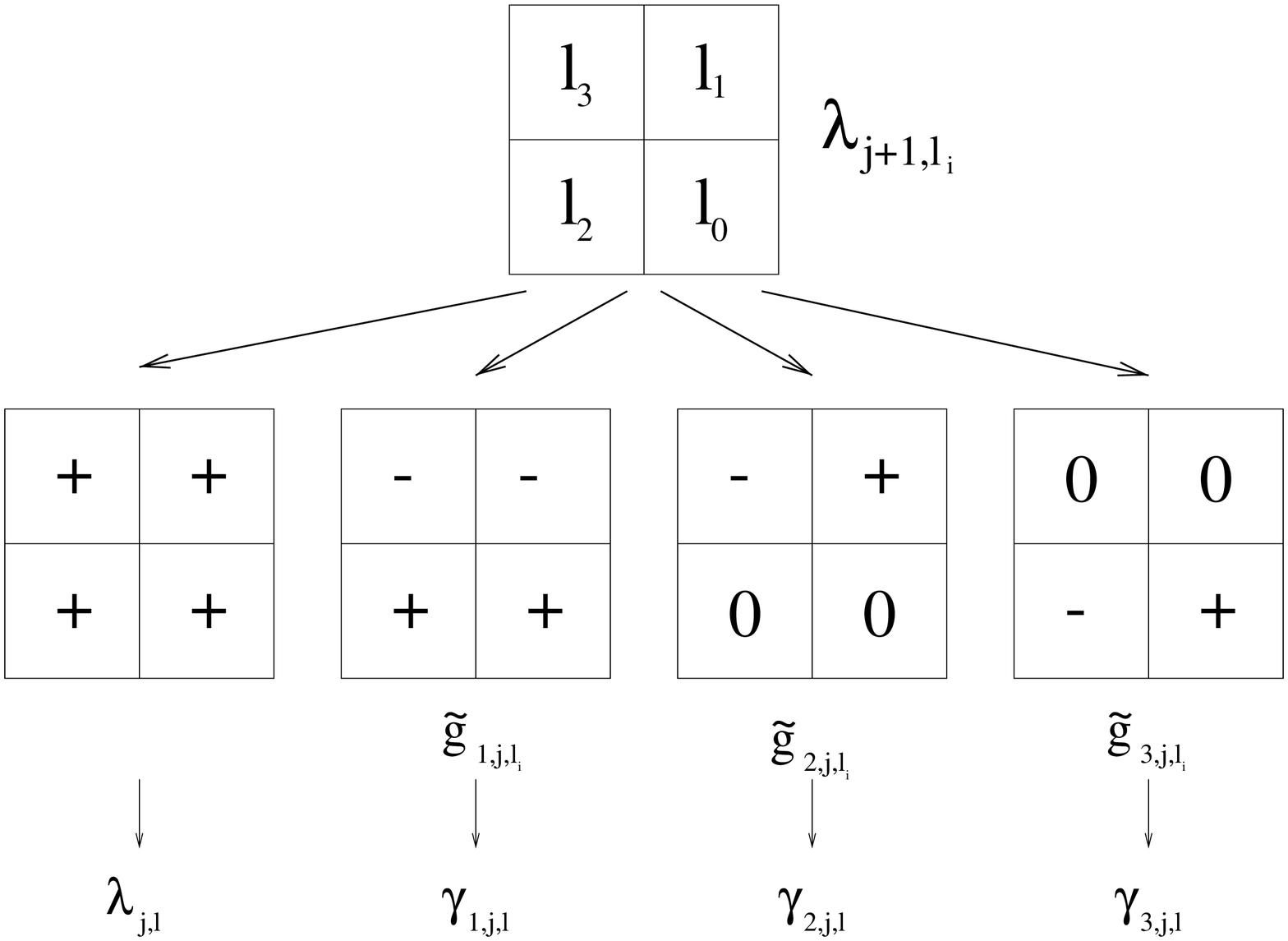,width=8cm}}
\caption{The procedure to obtain the wavelet coefficients at a
resolution level $j$ is shown. Each coefficient is obtained from a
linear combination of four approximation coefficients at the level
$j+1$. The new $\lambda_j,l$ are just the average over the four
original ones. The value of the $\tilde{g}_{m,j,l_i}$ factors
used to obtain the details is $+\mu_{j+1}$, $-\mu_{j+1}$ or $0$ 
as indicated in the figure.}
\label{gcoef}
\end{figure}
The temperature field $\Delta T/T$ on the sphere can then be written as
\begin{eqnarray}
\frac{\Delta T}{T}(x_i)&=&\sum_{l=0}^{n_{j_0}-1}
\lambda_{j_0,l}\varphi_{j_0,l}(x_i)\nonumber \\
& & + \sum_{m=1}^3\sum_{j=j_0}^{J-1}
\sum_{l=0}^{n_j-1}\gamma_{m,j,l}\psi_{m,j,l}(x_i)
\end{eqnarray}
where $\lambda_{j_0,l}$ and $\gamma_{m,j,l}$ are the approximation and
detail wavelet coefficients respectively.
The index $j$ runs over the different scales with $J$ the resolution of 
the original map and $j_0$ the coarsest resolution considered.

The procedure to obtain the wavelet coefficients is as follows 
(see Fig~\ref{cobeshw}).
We start with a map at a resolution $J$, identifying each pixel
with the $\lambda_{J,l}$ coefficients.
Each approximation and detail coefficient at position $l$ and
resolution level $J-1$
is obtained as a linear combination of the four corresponding pixels
$l_0$,$l_1$,$l_2$,$l_3$ at resolution $J$:
\begin{eqnarray}
\lambda_{J-1,l} & = & \frac{1}{4}\sum_{i=0}^{3}\lambda_{J,l_i} \\
\gamma_{m,J-1,l} & = & \sum_{i=0}^3\tilde{g}_{m,J-1,l_i}\lambda_{J,l_i}
\end{eqnarray}
The approximation coefficients at the new
resolution level are just the average over the four original coefficients
and, therefore, correspond to a lower resolution
version of the initial image. Complementary to those, the details 
encode the difference between the degraded and the original maps.
The $\tilde{g}_{m,j,l}$ coefficients take values $+\mu_{j+1}$,
$-\mu_{j-1}$ or $0$,
depending on the indices, as shown in Fig.~\ref{gcoef}. These
coefficients are obtained as an integral over the dual function
of $\psi_{m,j,l}$ in the corresponding support (see e.g. Tenorio et al. 1999).
Therefore, at level $J-1$ we have three different sets of details 
and one set of approximation coefficients, each of them 
with $n_{J-1}=n_J/4$ pixels.

Then, we apply the same procedure to the degraded map formed by
$\lambda_{J-1,l}$ , obtaining again a coarser
image and a set of details corresponding to the resolution level
$J-2$. Repeating this process down to the lowest resolution
considered in our analysis,
$j_0$, we are left with the details for resolutions $J-1$ to $j_0$
as well as the approximation $\lambda_{j_0}$ coefficients.
With this scheme it is also straightforward to identify those
coefficients which are descendants of pixels contaminated by the Galaxy
and discard them when performing the analysis.

To recover the map we just need to invert the process, starting with
the approximation and detail coefficients at the lowest resolution
$j_0$ up to the map at the initial resolution $J$.


\end{document}